\documentclass[prb, reprint,amsmath,amssymb,floatfix,showpacs]{revtex4-1}

\usepackage[colorlinks,citecolor=blue,linkcolor=blue]{hyperref}
\usepackage{graphicx}
\usepackage{dcolumn}
\usepackage{bm}
\usepackage{txfonts}

\DeclareMathOperator{\sinc}{sinc}

\begin{document}

%
%
\title{Breathing modes of confined skyrmions in ultrathin magnetic dots}

\author{Joo-Von Kim}
\email{joo-von.kim@u-psud.fr}
\author{Felipe Garcia-Sanchez}
\affiliation{Institut d'Electronique Fondamentale, UMR CNRS 8622, Univ. Paris-Sud, 91405 Orsay, France}
\author{Jo{\~a}o Sampaio}
\author{Constance Moreau-Luchaire}
\author{Vincent Cros}
\author{Albert Fert}
\affiliation{Unit{\'e} Mixte de Physique CNRS/Thales, 1 avenue Augustin Fresnel, 91767 Palaiseau, France}

\date{\today}

\begin{abstract}
The dynamics of individual magnetic skyrmions confined in ultrathin film dots is studied theoretically. The systems considered are transition metal ferromagnets possessing perpendicular magnetic anisotropy and particular attention is given to the dynamic response of the skyrmions to perpendicular driving fields. By using micromagnetics simulations, it is shown that breathing modes can hybridize with geometrically-quantized spin wave eigenmodes of the circular dots considered, leading to distinct features in the power spectrum that differ to the behavior expected for uniformly magnetized systems. The field dependence of the breathing modes offers a direct means of detecting and characterizing such skyrmion states in experiment.
\end{abstract}

\pacs{75.30.Ds, 75.40.Gb, 75.75.-c, 75.78.Fg}

\maketitle

\section{Introduction}
In thin film magnetism, the skyrmion represents a topological magnetic soliton that is stabilized by the presence of chiral magnetic interaction of the Dzyaloshinskii-Moriya form.~\cite{Dzyaloshinsky:1958vq, Moriya:1960go, Moriya:1960kc} Skyrmion states were predicted to appear in ultrathin films~\cite{Bogdanov:1989vt, Bogdanov:1999df, Rossler:2006cq} and their observation in a variety of physical systems~\cite{Muhlbauer:2009bc, Pappas:2009bk, Yu:2010iu, Heinze:2011ic, Seki:2012ie, Huang:2012ij, Kanazawa:2012gz, Milde:2013hd} has led to intense efforts to understand the physical conditions under which they are stable, along with their static and dynamic properties. Their topological nature has important consequences on electron and heat transport,~\cite{Neubauer:2009en, Shiomi:2013fz} which lead to novel features involving current-driven dynamics of skyrmion lattices.~\cite{Jonietz:2010fy, Everschor:2011cc, Schulz:2012bi} A recent demonstration of the controlled nucleation and annihilation of isolated skyrmions~\cite{Romming:2013iq} is promising for future applications exploiting these objects for information storage and processing.

In the context of potential spintronic applications using skyrmions, recent attention has turned toward ultrathin transition metal ferromagnets in multilayer structures lacking inversion symmetry, which are thought to harbor a sizeable Dzyaloshinskii-Moriya interaction (DMI). These structures exhibit a perpendicular magnetic anisotropy and comprise a heavy metal underlayer possessing strong spin-orbit coupling, on top of which the thin ferromagnetic metal is grown and capped by nonmagnetic layer. The DMI in such systems is argued to arise from an RKKY-type interaction that is mediated by the interface atoms of the heavy-metal underlayer through the spin-orbit interaction.~\cite{Fert:1980hr, Fert:1990, Crepieux:1998ux, Fert:2013fq} Some examples of such material systems include Pt/Co (0.6)/Al$_2$O$_3$,~\cite{Thiaville:2012ia} Pt (2.5)/Co (0.3)/Pt (1.5),~\cite{Je:2013es} Pt (3)/CoFe (0.6)/MgO (1.8) and  Ta (5)/CoFe (0.6)/MgO (1.8),~\cite{Emori:2013cl} Pt (1.5)/Co (0.3)/Ni (0.7)/Co (0.15),~\cite{Ryu:2013dl}, and Pt/[Ni/Co]$_n$ and Ir/[Ni/Co]$_n$,~\cite{Chen:2013bc} where the figures in parentheses represent film thicknesses in nm and are presented here to highlight the importance of the interfacial origin of the effect. At present, the existence of a sizeable DMI has been inferred from experiments involving imaging of static domain wall structures,~\cite{Chen:2013bc} domain wall nucleation,~\cite{Pizzini:2014wq} and from field~\cite{Je:2013es} and current-driven magnetic domain wall dynamics~\cite{Emori:2013cl, Ryu:2013dl} in which a clear dependence on the wall chirality has been observed. In these cases, the experimental results suggest that the domain wall structure involves a N{\'e}el profile, rather than a Bloch form, which is possible in these strong ferromagnets if a sufficiently large DMI is present.~\cite{Heide:2008da, Thiaville:2012ia} From the strength of the DMI obtained in these experiments, theoretical studies suggest that isolated skyrmions should at least be metastable in these systems.~\cite{Sampaio:2013kn}

In this article, we explore the dynamics of individual skyrmions confined geometrically in ultrathin circular dots. In particular, the breathing dynamics of the skyrmion is explored, which is known to provide a distinct microwave response from experiments conduced on skyrmion crystals in helimagnetic insulators.~\cite{Onose:2012bc} Particular attention is given to how the breathing dynamics is affected by the confinement and to the interaction between the breathing mode and the geometrically-quantized spin wave eigenmodes of the dot. The dynamics is studied primarily in static perpendicular applied fields with driving ac perpendicular fields, where it is shown that the magnetic response in this geometry provides clear signatures of the skyrmion excitation and therefore an alternative means of quantifying the DMI.

This article is organized as follows. In Section II, a description of the model and simulation method is given. In Section III, the static properties of isolated skyrmions in circular dots is described. The dynamic response of these skyrmions to driving perpendicular fields is presented in Section IV. Some discussion and concluding remarks are given in Section V.

\section{Simulation Model and Method}
The static and dynamic states of the isolated skyrmion in a confined magnetic dot were studied using numerical micromagnetics simulations with a modified version of the \textsc{mumax2} code.~\cite{Vansteenkiste:2011bx} The code performs a numerical time integration of the Landau-Lifshitz equation of motion with Gilbert damping for the magnetization dynamics, 
\begin{equation}
\frac{d \mathbf{m}}{dt} = -|\gamma_0| \mathbf{m} \times \mathbf{H}_{\rm eff} + \alpha \mathbf{m} \times \frac{d \mathbf{m}}{dt},
\end{equation}
where $\gamma_0$ is the gyromagnetic constant, $\alpha$ is the phenomenological Gilbert damping constant, $\| \mathbf{m} \|  =1$ is the unit vector representing the orientation of the magnetization, and $\mathbf{H}_{\rm eff}$ is the effective field,
\begin{equation}
\mathbf{H}_{\rm eff} = -\frac{1}{\mu_0 M_s} \frac{\delta U}{\delta \mathbf{m}},
\end{equation}
which represents the variational derivative of the total micromagnetic energy $U$ with respect to the magnetization. $M_s$ is the saturation magnetization. In addition to the usual Zeeman, isotropic exchange  (characterized by the exchange constant $A$) and dipole-dipole interactions, the total energy also includes a uniaxial anisotropy perpendicular to the film plane, $U_K = -K_u \int dV\; m_z^2 $, and an interfacial Dzyaloshinskii-Moriya interaction of the form
\begin{equation}
U_{\rm DM} = D \int dV \; \left[ m_z \left(\mathbf{\nabla} \cdot \mathbf{m}\right) - \left( \mathbf{m} \cdot \mathbf{\nabla} \right) m_z \right],
\label{eq:DMI}
\end{equation}
which is consistent with the induced coupling associated with the strong spin-orbit underlayer in the ultrathin film geometry considered here.~\cite{Bogdanov:2001hr, Thiaville:2012ia, Fert:2013fq, Rohart:2013ef, Sampaio:2013kn} The strength of the Dzyaloshinskii-Moriya interaction is characterized by the constant $D$ and its sign determines the preferred handedness of the chiral micromagnetic ground state. For the simulation results presented here, the system was discretized using rectangular finite difference cells $1.5625 \times 1.5625 \times d$ nm in size, corresponding to a $d=1$ nm thick film. The micromagnetic parameters used correspond to a model thin film perpendicular anisotropy system: an isotropic exchange constant of $A = 15$ pJ/m, a perpendicular anisotropy constant of $K_u = 1$ MJ/m$^3$, and a saturation magnetization of $M_s = 1$ MA/m.~\cite{GarciaSanchez:2014}

As discussed in the introduction, the main focus of this work concerns the breathing response of the individual skyrmion in a dot. To this end, the calculations performed involve the dynamic response of the perpendicular magnetization to perpendicular dynamic fields $h_z$,
\begin{equation}
m_z(\omega) = \chi_{zz}(\omega) h_z(\omega).
\end{equation}
In the absence of the DMI, the equilibrium ground state is uniformly magnetized along the easy $z$ axis, which means that the linear susceptibility $\chi_{zz}$ is vanishingly small. In the context of spin wave excitation, this corresponds to the parallel pumping geometry in which the lowest-order response involves nonlinear spin wave processes that appear above a finite threshold of the pumping field.~\cite{Sparks:1964vb} However, the presence of the DMI in a finite-sized system necessarily leads to a nonuniform magnetic configuration. First, magnetization tilts away from the easy axis at boundary edges as a result of nontrivial boundary conditions, leading to a state that can be described by pinned partial N{\'e}el domain walls at the edges.~\cite{Rohart:2013ef, GarciaSanchez:2014} Note that these tilts exist even if the sample is uniformly magnetized away from the edges. Second, a skyrmion involves large magnetization tilts across its core, with variations between $m_z \pm 1$ that can extend over a sizeable fraction of the dots studied, as shown in the following. These tilts allow for coupling to the perpendicular driving field and lead to a finite response for $\chi_{zz}$ even for vanishingly small fields $h_z$.  Furthermore, the use of perpendicular driving fields minimizes the complication involved with the skyrmion gyration,~\cite{Moon:2014cs} which is known to be complex in such confined geometries. Translational motion of the skyrmion core can also become complicated due to scattering from the dot edges~\cite{Iwasaki:2013hb} and is minimized in the field geometry considered. Therefore, a finite $\chi_{zz}$ in the linear response regime should provide a clear signature of individual skyrmions in magnetic dots.

Calculation of the dynamic response in our simulations involves two steps. The ground state of the magnetic dot with an individual skyrmion is first determined in the presence of a magnetic field applied perpendicular to the film plane at zero temperature. The initial state comprises an hedgehog skyrmion whose core profile is computed using an \emph{ansatz} for the radial variation of the polar angle of the magnetization vector.~\cite{Kiselev:2011cm} The in-plane components of the magnetization point along the radially with respect to the core center, which is the preferred N{\'e}el-like state associated with the DMI assumed in Eq.~\ref{eq:DMI}.~\footnote{This method is general and can be applied to other forms of the DMI, namely for vortex-like states in MnSi and FeGe} For each value of the applied magnetic field, this initial state is then allowed to relax over 5 ns with a large damping constant of $\alpha = 0.3$. The time-variation of the magnetization during this relaxation process is monitored to ensure that the micromagnetic state converges to an equilibrium state to the limit of the numerical precision of simulation. The magnetization dynamics is then computed about this equilibrium state. The excitation modes of the skyrmion are determined from the transient response of the system to a time-varying but spatially uniform field applied in the perpendicular $z$ direction. A cardinal sine function, $\sinc(t) = \sin(\pi \nu t)/(\pi \nu t)$, is used for the time-dependence of this field as its Fourier transform is a rectangular function. A cut-off frequency of $ \nu =$ 100 GHz was used for the cardinal sine function. The amplitude of the excitation field is 0.5 mT. The transient dynamics was computed over 20 ns, with data taken every 5 ps, using a smaller value of the damping constant, $\alpha = 0.002$, which allowed for better frequency resolution of the excited modes. The power spectral density of the skyrmion excitations is then computed using the transient response of the perpendicular component of the spatially averaged magnetization, $\langle m_z \rangle = (1/V) \int dV \; m_z(\mathbf{r})$.

\section{\label{sec:statics}Static properties of isolated skyrmions}
\begin{figure}
\centering\includegraphics[width=6.0cm]{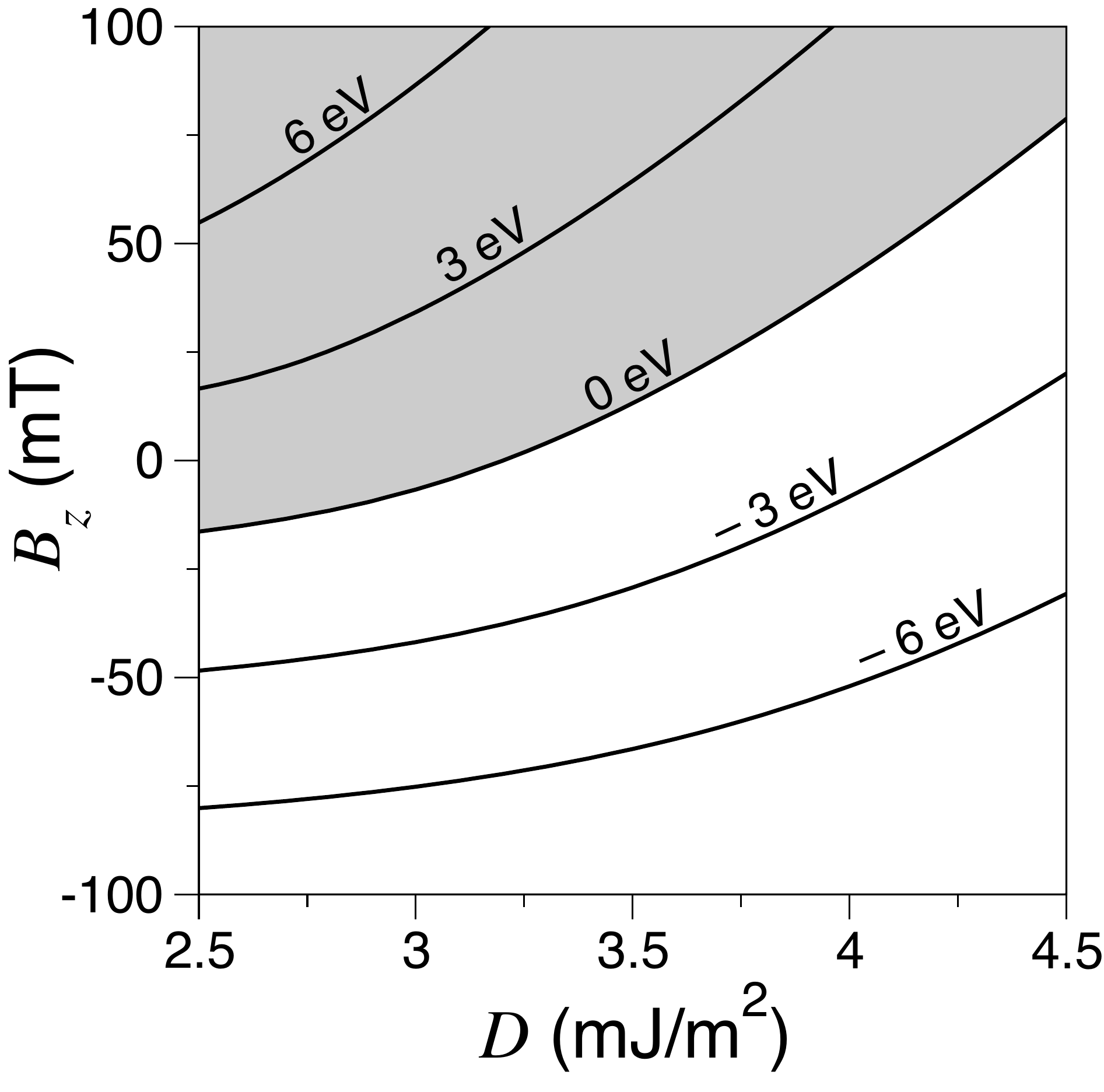}
\caption{Energy contours of $\Delta U$ related to the metastability of the isolated skyrmion state with respect to the quasi-uniform state. $\Delta U > 0$ indicates a lower energy for the quasi-uniform state, while $\Delta U < 0$ indicates a lower energy for the isolated skyrmion state.}
\label{fig:EnergyContours}
\end{figure}

The static properties of single skyrmion states in 100 nm diameter circular dots are described in this section. It has been shown from previous work that the skyrmion in a dot can represent a metastable state or a global energy minimum depending on the strength of the DMI.~\cite{Sampaio:2013kn} Here, the analysis is extended to include an applied magnetic field along the easy axis perpendicular to the field plane, which is known to modify the stability of the skyrmion state in extended systems.~\cite{Kiselev:2011cm} In Fig.~\ref{fig:EnergyContours}, the contours corresponding to the energy difference $\Delta U = U_{\rm sky} - U_{\rm uni}$ is shown.
$U_{\rm sky}$ corresponds to the total micromagnetic energy of the dot with a skyrmion state, while $U_{\rm uni}$ describes the energy of the state quasi-uniformly magnetized along $+z$ at the center of the dot with tilts at the dot edges. Both states are computed using the relaxation method described in the previous section. The zero energy contour in Fig.~\ref{fig:EnergyContours} therefore represents the values of $(B_z, D)$ for which both the skyrmion and uniform states possess the same energy. The region of the phase diagram in which the skyrmion is metastable is shown in gray and possesses positive energy contours, while the region in which the skyrmion is a global energy minimum state is shown in white and possesses negative energy contours.

\begin{figure}
\centering\includegraphics[width=8.0cm]{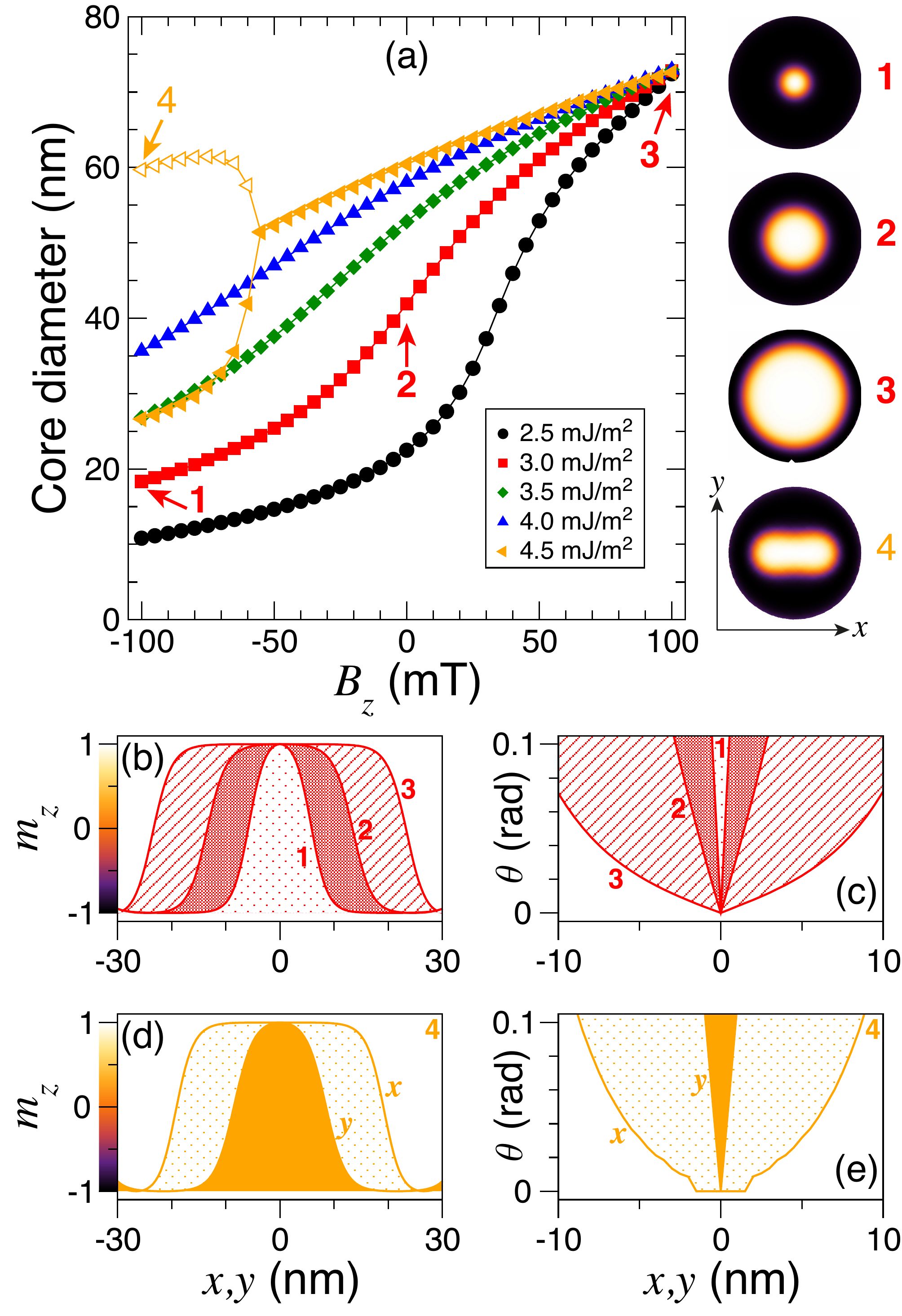}
\caption{(Color online) Static skyrmion profiles for four different cases. (a) Core diameter as a function of applied perpendicular field for varying DMI strength. The right inset depicts the static $m_z$ magnetization component. (b,d) Spatial profiles of $m_z$ across the skyrmion core. The color scale used in (a) is shown in the vertical axis of the plot. (c,e) Spatial profiles of the magnetization polar angle $\theta$ across the skyrmion core.}
\label{fig:radius_vs_Bz}
\end{figure}

The perpendicular field plays an important role in determining the size of the skyrmion core, as shown Fig.~\ref{fig:radius_vs_Bz}.
Here, the magnetization at the core center is oriented along the $+z$ direction, while magnetization outside the core is oriented in the opposite direction along $-z$. Four profiles are shown in the panel to the right of Fig.~\ref{fig:radius_vs_Bz}, where the perpendicular magnetization component $m_z$ is illustrated using the color scheme depicted along the vertical axes of Figs.~\ref{fig:radius_vs_Bz}b and \ref{fig:radius_vs_Bz}d. For the cases $D = $ 2.5 to 4.0 mJ/m$^2$ considered, the core remains circular under the range of applied fields used, and expands with increasing fields whilst contracting with decreasing fields. The profiles for $D = $ 3.0 mJ/m$^2$ are illustrated in Fig.~\ref{fig:radius_vs_Bz} at negative, zero, and positive applied fields and are labelled `1', `2', and `3', respectively. The expansion of the core is limited by the dot boundary, where the tilts in the magnetization exert an inward force on the skyrmion core thereby leading to a confining potential.~\cite{Sampaio:2013kn, Rohart:2013ef} As a result, the core diameter converges toward a limiting value at large positive applied fields for the different cases of $D$ considered, as illustrated by the state `3'. The core contracts in a monotonous fashion for decreasing and negative applied fields, except for the $D = 4.5$ mJ/m$^2$ case for which a change in shape of the core is seen. In contrast to the other cases in which the core remains circular, it assumes an elliptical shape for $D = 4.5$ mJ/m$^2$ below a certain critical negative field ($B_z = -55$ mT) as illustrated by the state `4'. This elongation of the core results from a competition between the Zeeman energy, which favors the contraction of the core such that regions of magnetization along $-z$ are maximized, and the DMI, which reduces the domain wall energy and favors expansion of the core to maximize its circumference. The two branches for state `4' for $B_z < -55$ mT in Fig.~\ref{fig:radius_vs_Bz}a correspond to the length and width of the core.

Cross-sections of the core profile corresponding to states `1' to `4' are shown in Figs.~\ref{fig:radius_vs_Bz}b-e. The compact nature of the core can be seen from the profile of the polar angle, $\theta = \cos^{-1}(m_z)$, which is shown in Fig.~\ref{fig:radius_vs_Bz}c for states `1', `2', and `3' and in Fig.~\ref{fig:radius_vs_Bz}e for state `4'. The angle $\theta$ is seen to vary linearly around the core center for states `1', `2', and `3', with a discontinuity in $\partial\theta/\partial (x,y)$ at the core center, which is consistent with ansatz described in Ref.~\onlinecite{Kiselev:2011cm}. It is interesting to note that the elliptical core of state `4' possesses an extended bubble-like character along its long axis, while a discontinuity in $\partial\theta/\partial y$ is observed at the core center.

\section{\label{sec:dynamics}Dynamic response to driving perpendicular fields}

\begin{figure}
\centering\includegraphics[width=8.6cm]{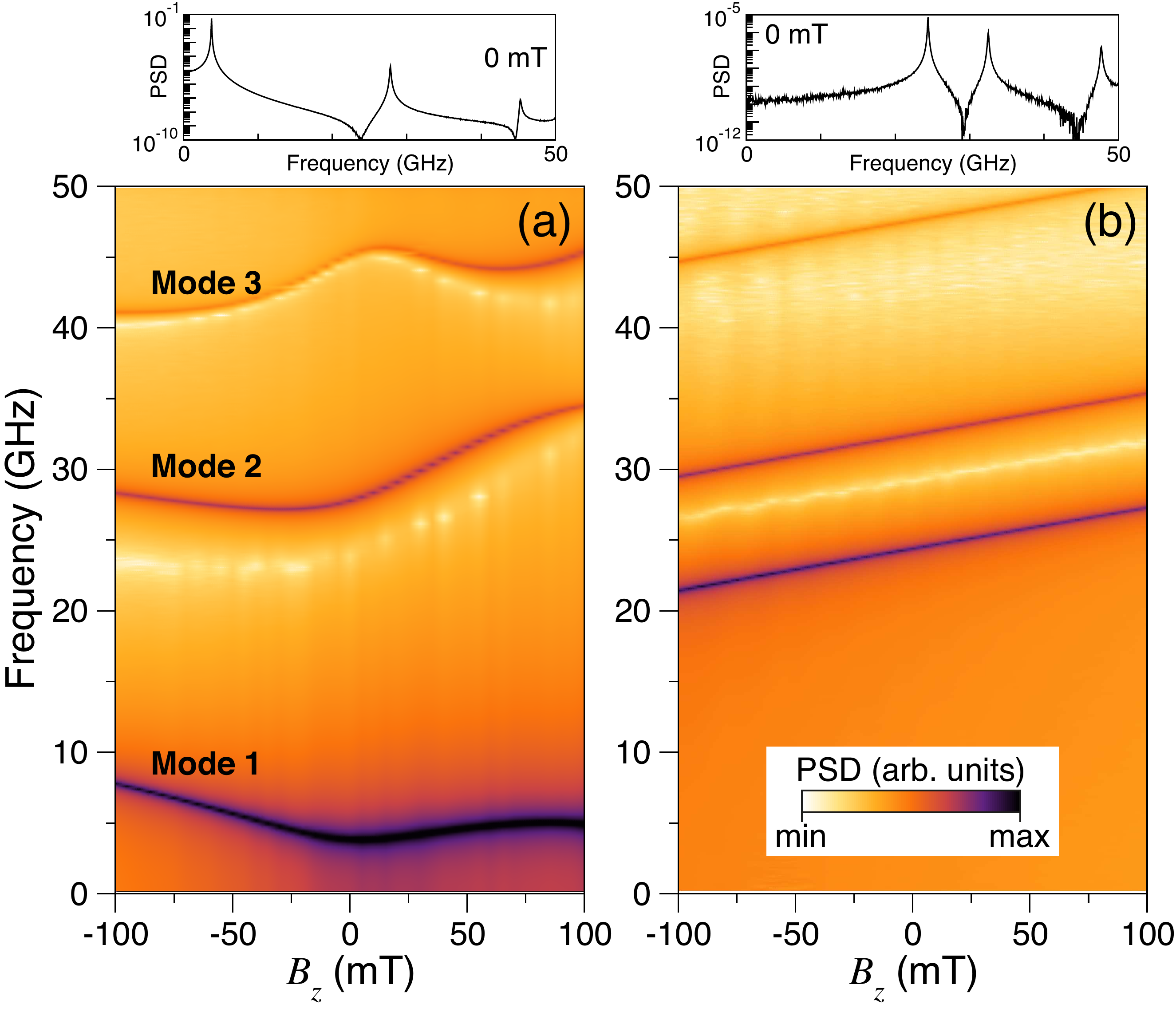}
\caption{(Color online) Map of the power spectral density (PSD) of excitations as function of perpendicular applied field $B_z$ for $D=3$ mJ/m$^2$. (a) Single isolated skyrmion in ground state. (b) Nominally uniform ground state. The insets at the top show the corresponding (PSD) at zero field.}
\label{fig:freq_vs_Bz_example}
\end{figure}

The dynamic properties of the single skyrmion states in 100 nm diameter circular dots are described in this section. An example of the calculated power spectrum of excitations is shown in Fig.~\ref{fig:freq_vs_Bz_example} for an intermediate value of the DMI, $D = 3$ mJ/m$^2$.
Fluctuations in the perpendicular component of the magnetization are considered, $\delta m_z(t) = \langle m_{z,0} \rangle - \langle m_z(t) \rangle$, where $m_{z,0}$ corresponds to the equilibrium state and the angular brackets represent a spatial average of the magnetization, as discussed previously. The average magnetization is used here for the analysis as this quantity would be the most accessible component in experiment. The power spectrum, $\mathcal{S}(\omega)$, is computed from the Fourier transform of this fluctuation,
\begin{equation}
\mathcal{S}(\nu) = \left| \int_{0}^{t_{0}} dt \; e^{-i 2 \pi \nu t} \delta m_z(t) \right|^2,
\end{equation}
where $t_0$ represents the total duration of the time integration performed in the simulations. A color map of this function is presented in Fig.~\ref{fig:freq_vs_Bz_example}, where the inset shows a plot of this function at zero field in the frequency range $\nu = 0$  to 50 GHz. In Fig.~\ref{fig:freq_vs_Bz_example}a, the ground state comprises an individual skyrmion at the center of the dot, as shown in Fig.~\ref{fig:radius_vs_Bz}. Three distinct modes are observed in the frequency range considered, whose frequencies exhibit a nontrivial dependence on the applied magnetic field. To highlight the distinct features associated with the presence of the skyrmion, the power spectrum for a dot with no skyrmion is shown in Fig.~\ref{fig:freq_vs_Bz_example}b. In this case, the magnetization is uniformly magnetized in the $+z$ direction at the center of the dot, but it tilts at the edges as a result of the boundary conditions related to the DMI.~\cite{Rohart:2013ef, GarciaSanchez:2014} Because of these tilts,  $\chi_{zz}$ is finite even for this case and three modes can be seen in the same frequency range. The lowest mode corresponds to the uniform precession mode and all three modes exhibit a linear field dependence, which is expected for spin waves in this geometry. It is interesting to note that the lowest-order mode in the skyrmion case exhibits a much lower frequency than the FMR mode in the uniform case, while the second and third order modes have similar frequencies. The FMR mode is also absent in the skyrmion case. Together, these features suggest that the appearance of a low-frequency excitation, such as the branch below 10 GHz in Fig.~\ref{fig:freq_vs_Bz_example}a, is a strong signature of the presence of a confined skyrmion in a dot and a measure of the DMI.

\begin{figure}
\centering\includegraphics[width=8.0cm]{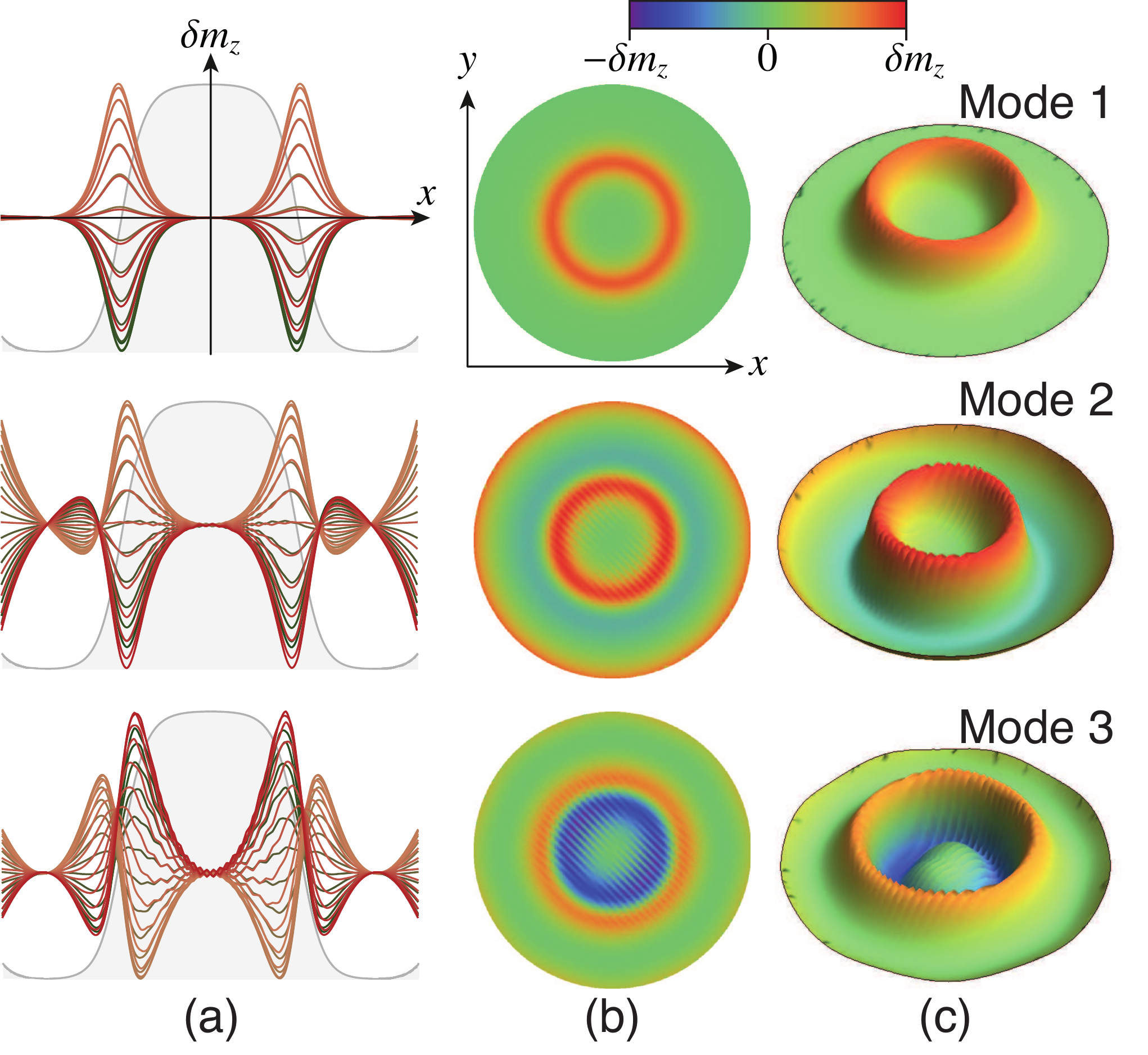}
\caption{(Color online) Spatial profiles the fluctuation in the magnetization component $\delta m_z$ of the skyrmion breathing modes for $D = 3$ mJ/m$^2$ at $B_z = 0$ mT. (a) Spatial profile of the fluctuation across the dot, where each curve represents a different instant $t$ over one period of oscillation with the static core profile given in grey. (b) A two-dimensional snapshot of the mode. (c) A three-dimensional snapshot of the mode.}
\label{fig:ModeProfiles01}
\end{figure}

Representative spatial profiles of the three confined skyrmion modes in Fig.~\ref{fig:freq_vs_Bz_example}a are presented in Fig.~\ref{fig:ModeProfiles01} for zero applied static field ($B_z = 0$).
Each profile has been calculated with a ``stroboscopic'' method as follows. The mode frequency $\nu_k$ is first determined from a Lorentzian fit to the power spectrum under the field condition studied (e.g., inset in Fig.~\ref{fig:freq_vs_Bz_example}a). The dynamics is then simulated with a spatially-uniform sinusoidal magnetic field applied along the $+z$ axis, $b_z(t) = b_0 \sin(2 \pi \nu_k t)$ with $b_0 = 0.5$ mT, over 20 periods of the excitation with the precomputed equilibrium state used as the initial state in the simulations. The micromagnetic state is then saved at time intervals of $1/25\nu_k$ over the last 10 periods of the simulations, which leads to 10 snapshots of the state for each instant $t_i = 0, 1/25\nu_k, 2/25 \nu_k, \ldots, 23/25\nu_k, 24/25\nu_k$ relative to the phase of the field excitation. The final mode profile at each $t_i$ is then obtained by averaging over the 10 snapshots, which allows features related to non-resonant dynamics and numerical artifacts to be averaged out over a single period of excitation. Each row in Fig.~\ref{fig:ModeProfiles01} corresponds to each of the three branches seen in Fig.~\ref{fig:freq_vs_Bz_example}a, where the mode index is given in the right column (Fig.~\ref{fig:ModeProfiles01}c). In the left column (Fig.~\ref{fig:ModeProfiles01}a), the spatial profile of the fluctuation $\delta m_z(x)$ across the $x$-axis of the dot is shown, where each curve corresponds to a successive instant, with a time increment of $1/25 \nu$, over one period of oscillation, $1/\nu$. The static skyrmion profile $m_{z,0}(x)$ is shown in the background as a filled grey curve, where the full variation between $m_{z,0} = \pm 1$ is shown (note that this is not the same scale used for $\delta m_z(x)$ on the same graph). In the center column (Fig.~\ref{fig:ModeProfiles01}b), the spatial profile of the fluctuation across the dot, $\delta m_z(x,y)$, is shown at one instant as a two-dimensional color plot, where the color code used is given at the top of the figure. The same profile is rendered as a three-dimensional plot in the right column (Fig.~\ref{fig:ModeProfiles01}c), where the corresponding mode amplitude is shown along with the color code used.

From these images, mode 1 can be deduced to be the breathing mode of the skyrmion, where the magnetization fluctuation is localized to the core region of the skyrmion. From the time dependence in Fig.~\ref{fig:ModeProfiles01}a, the breathing at opposite sides of the core is seen to be in phase, which indicates that the core shrinks and expands whilst preserving its radial symmetry with no discernible azimuthal component to the dynamics. Mode 2 corresponds to a hybridization between the breathing mode and a radial spin wave mode of the circular dot. The amplitude of this excitation is largest around the core, but a significant component also appears at the dot edges. Mode 3 is a similar hybridized mode, with the key difference being the larger mode amplitude occurring within the core itself and one node in the mode profile appearing at the core boundary. Its profile in Fig.~\ref{fig:ModeProfiles01}a shows that the mode excitation is largely confined to the core, with weaker coupling to the edge magnetization compared to mode 2. For modes 2 and 3, the radial symmetry of the equilibrium skyrmion profile is preserved by the excitation field in a similar way to mode 1.

\begin{figure}
\centering\includegraphics[width=8.6cm]{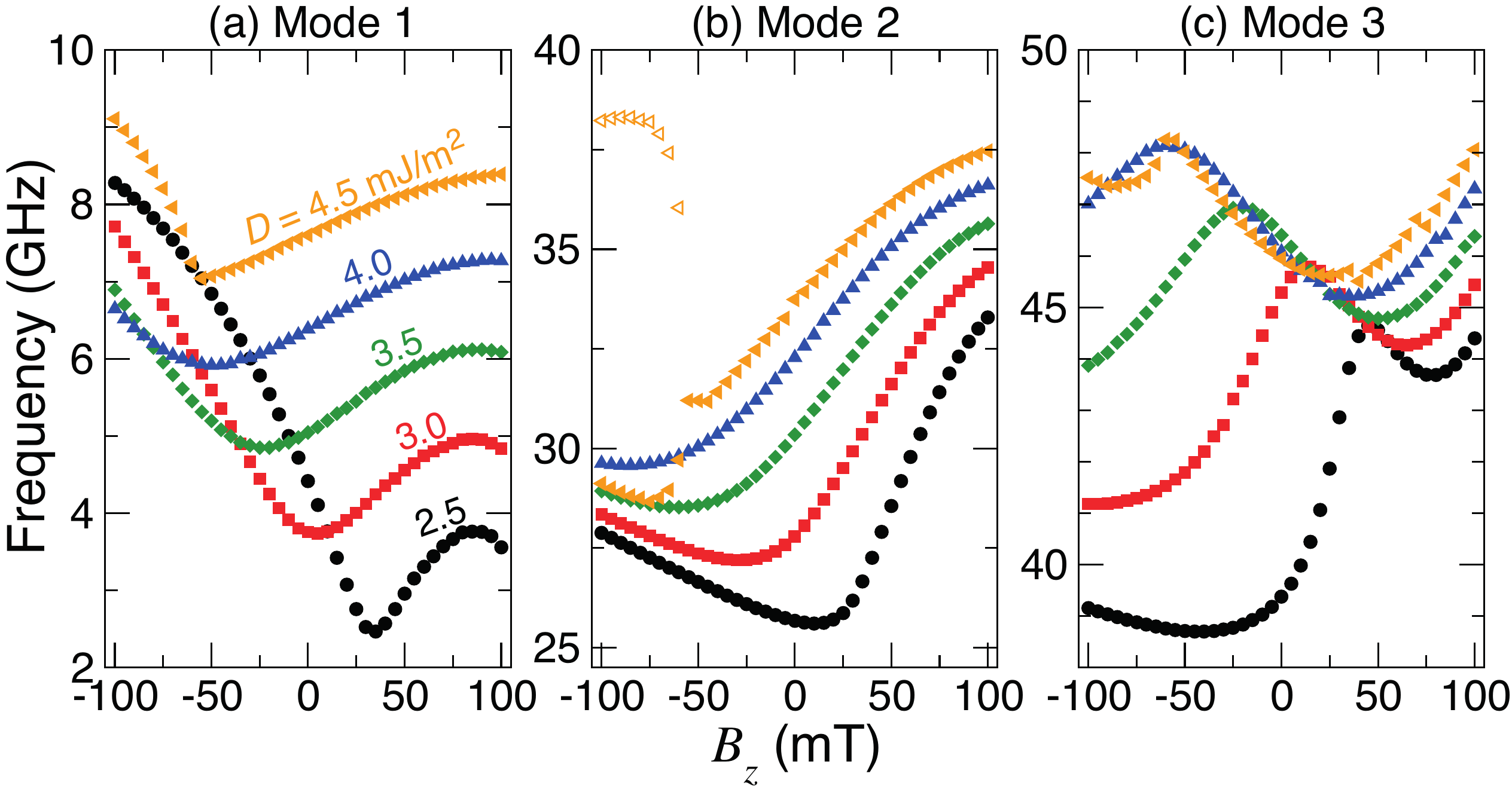}
\caption{(Color online) Mode frequency as a function of perpendicular applied field $B_z$ for different strengths of the DMI, $D$. (a) Mode 3. (b) Mode 2, with a frequency splitting seen for $D = 4.5$ mJ/m$^2$ in the field region $B_z \leq - 60 $ mT. (c) Mode 1. }
\label{fig:freq_vs_Bz_modes}
\end{figure}

The field dependence of the mode frequencies of the confined skyrmion is shown in Fig.~\ref{fig:freq_vs_Bz_modes}.
The results are presented for the three lowest order modes and for several values of the DMI strength. For the different cases considered, the frequency of each mode exhibits a similar variation with the applied static field. For mode 1 (Fig.~\ref{fig:freq_vs_Bz_modes}c), a minimum in the frequency is observed (i.e., near $B_z = 35$ mT for $D = 2.5$ mJ/m$^2$), the position of which shifts toward negative fields as the strength of the DMI is increased. The position of this frequency minimum coincides with the applied field value at which variations in the core diameter are greatest with the applied field. This is most pronounced for the $D = 2.5$ mJ/m$^2$ case, where the variation of the core diameter with $B_z$ exhibits the largest slope around $B_z = 35$ mT (Fig.~\ref{fig:radius_vs_Bz}a). These two features are consistent, since the frequency minimum reflects the fact that the breathing mode of the skyrmion is the least energetic when its core size can change the most readily with the applied magnetic field. Distinct features associated with this field also appear in the frequencies of modes 2 and 3, where a local extremum occurs.

\begin{figure}
\centering\includegraphics[width=8.0cm]{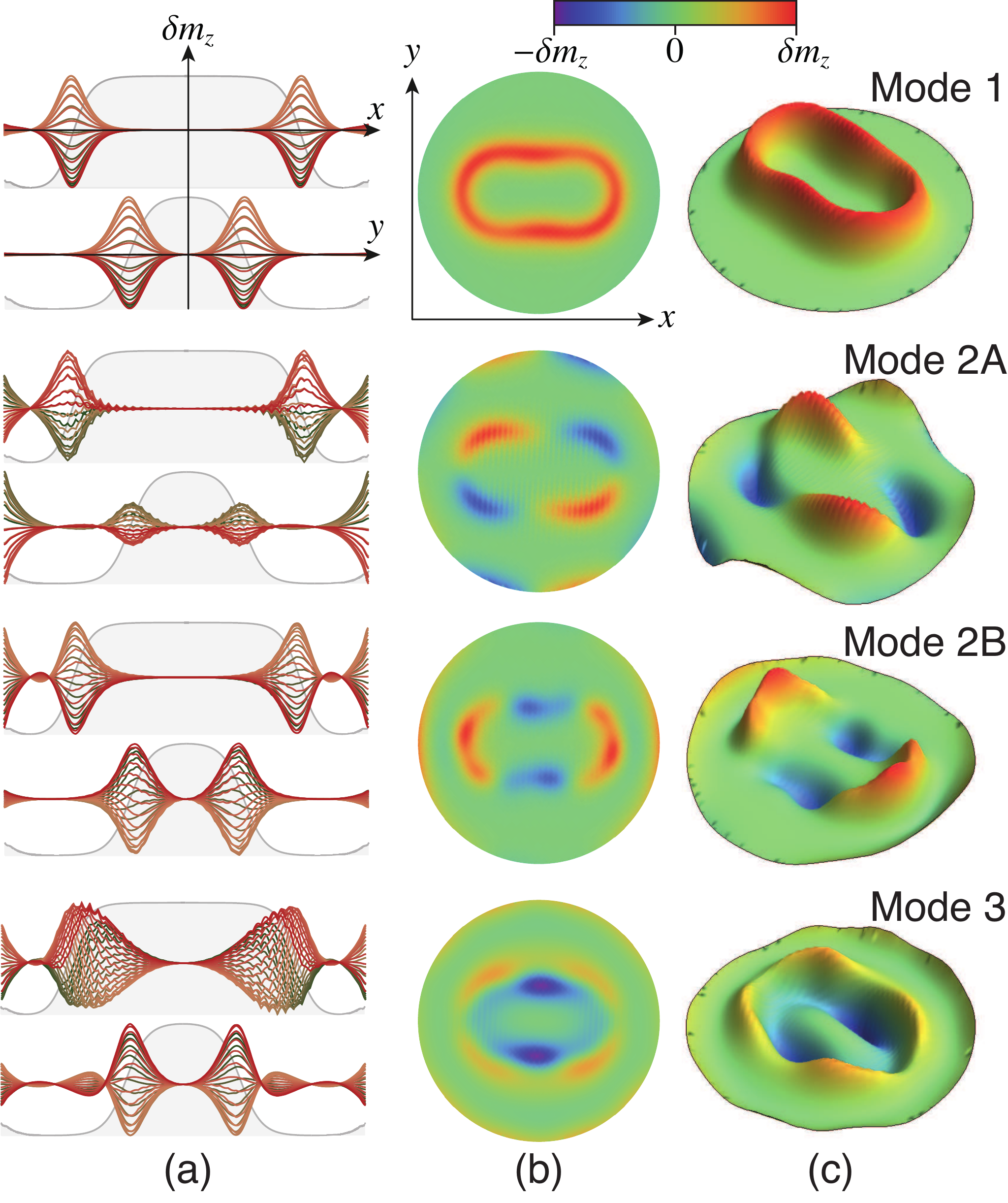}
\caption{(Color online) Spatial profiles of the skyrmion breathing modes for $D = 4.5$ mJ/m$^2$ at $B_z = -75$ mT. (a) Spatial profile of the fluctuation across the dot, where each curve represents a different instant $t$ over one period of oscillation with the static core profile given in grey. (b) A two-dimensional snapshot of the mode. (c) A three-dimensional snapshot of the mode. }
\label{fig:ModeProfiles03}
\end{figure}

One standout feature involves the splitting of mode 2 for $D = 4.5$ mJ/m$^2$ at negative fields $B_z < -55$ mT (Fig.~\ref{fig:freq_vs_Bz_modes}b). This splitting is associated with the transition from a circular to an elliptical shape for the skyrmion core, as discussed in Section~\ref{sec:statics} (Fig.~\ref{fig:radius_vs_Bz}). The magnitude of the splitting is also significant, with a difference of 9.1 GHz in frequency between modes 2A and 2B at $B_z = -100$ mT, which is comparable to the frequency of mode 1 at this field. The mode profiles at $B_z = -75$ mT are shown in Fig~\ref{fig:ModeProfiles03}.
While mode 1 retains its breathing character, which can be seen by the oscillations of $\delta m_z$ along the long and short axes of the core that remain in phase, the split modes 2A and 2B exhibit oscillations that are in anti-phase along the long and short axes (Fig.~\ref{fig:ModeProfiles03}a). This is characteristic of the presence of an azimuthal component to the breathing mode, which is observed in the two- and three-dimensional snapshots of the spatial profile in Figs.~\ref{fig:ModeProfiles03}b and ~\ref{fig:ModeProfiles03}c. While both 2A and 2B possess similar profiles, they differ in how the edge magnetization states are affected; for mode 2A, the edge magnetization precession amplitude is largest along the short axis of the core, while for mode 2B the amplitude is largest along the long axis of the core. An azimuthal component also appears for mode 3, where the internal structure of the breathing mode shows differences along the short and long axes of the core. In contrast to mode 2, however, no frequency splitting is seen for mode 3 as a result of the azimuthal component of the internal mode.

\begin{figure}
\centering\includegraphics[width=6.0cm]{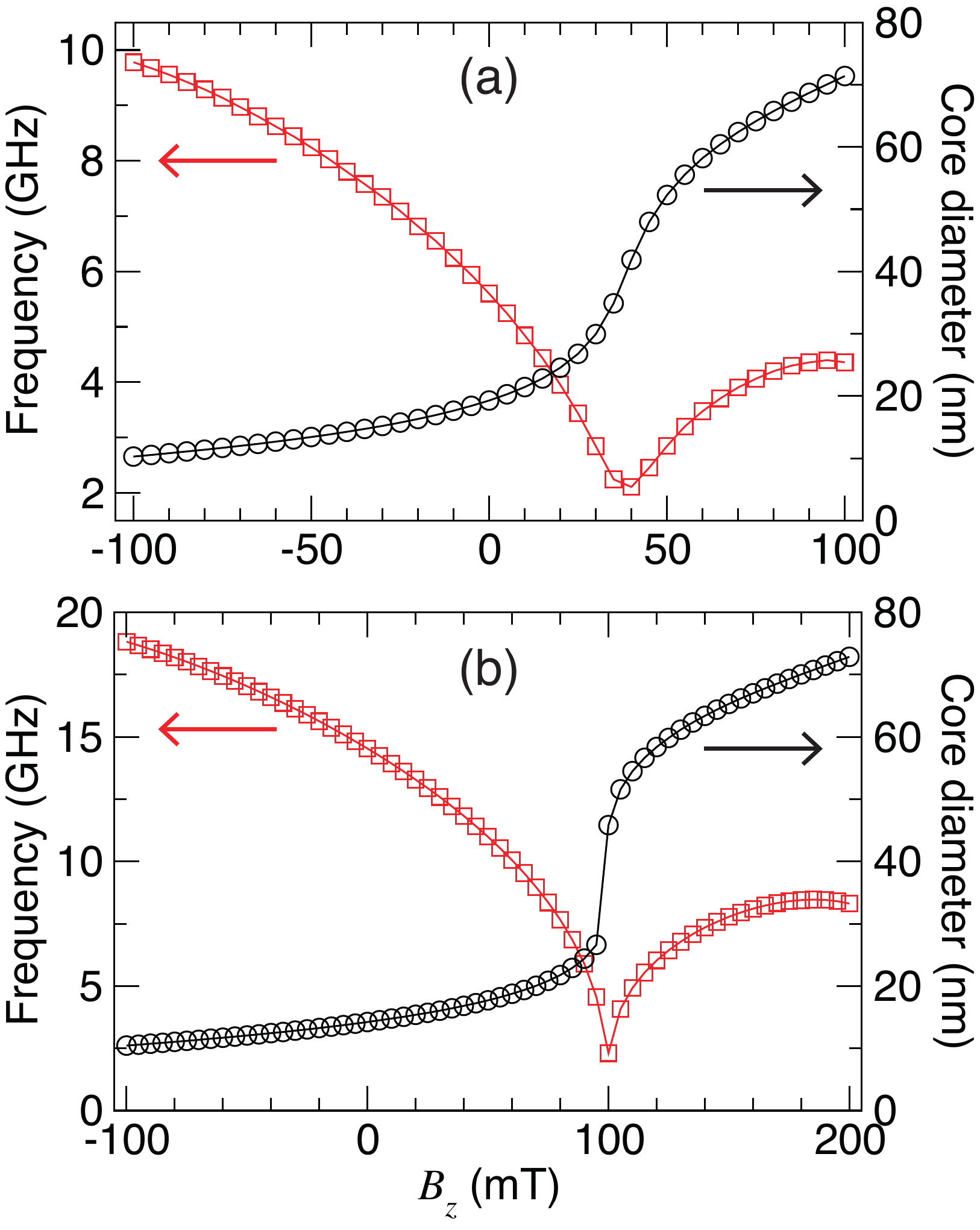}
\caption{(Color online) Variation of the breathing mode frequency (squares) and skyrmion core diameter (circles) as a function of perpendicular applied field, $B_z$. (a) Micromagnetic parameters from Thiaville \emph{et al}. (Ref.~\onlinecite{Thiaville:2012ia}). (b) Micromagnetic parameters from Sampaio \emph{et al}. (Ref.~\onlinecite{Sampaio:2013kn}).}
\label{fig:freq_vs_Bz_BC}
\end{figure}

The simulation results presented above have been conducted using parameters for a model system. While this set does not correspond to a particular material system \emph{per se}, it is representative of the class of ultrathin ferromagnetic films with perpendicular magnetic anisotropy as described in the introduction. As such, the qualitative features of the mode dynamics are expected to apply to systems that possess similar micromagnetic parameters. To illustrate one aspect of the universality of some of the features presented, the perpendicular field dependence of the breathing mode frequency and core radius for parameters used in other studies pertaining to chiral domain wall dynamics, as studied by Thiaville \emph{et al}. ($A = 16$ pJ/m, $K_u = 1.27$ MJ/m$^3$, $M_s = 1.1$ MA/m, $d = 0.6$ nm),~\cite{Thiaville:2012ia} and skyrmion dynamics, as studied by Sampaio \emph{et al}. ($A = 15$ pJ/m, $K_u = 0.8$ MJ/m$^3$, $M_s = 0.58$ MA/m, $d = 0.4$ nm),~\cite{Sampaio:2013kn} are shown in Fig.~\ref{fig:freq_vs_Bz_BC}.
The key differences in these parameter sets relate to the saturation magnetization and the film thickness, which result in different dipolar fields experienced by the skyrmion core. Nevertheless, the coincidence of the minimum in the breathing mode frequency with the largest susceptibility in the core diameter is remains a common feature, which is very pronounced in Fig.~\ref{fig:freq_vs_Bz_BC}b. Note that the frequency range over which the breathing mode varies is markedly larger for the parameters  of Sampaio \emph{et al}. for comparable changes in the core diameter. This is due to the larger exchange stiffness constant for this parameter set, $\mathcal{D} = \gamma A / M_s$, by virtue of the lower value of the saturation magnetization used. The exchange stiffness constant governs the quadratic dispersion relation of spin waves in the exchange dominated regime, $\omega(k) \sim \mathcal{D} k^2$. It plays a role in determining the mode spacing and the frequency range over which the breathing mode occurs.

\section{Discussion and Concluding Remarks}
Previous work on Fe/Ir(111) layers have shown that the magnitude of the DMI between two Fe atoms is approximately 1.8 meV.~\cite{Heinze:2011ic} As discussed by Sampaio \emph{et al}.,~\cite{Sampaio:2013kn} this atomistic interaction translates to a Dzyaloshinskii-Moriya constant of $D \simeq$ 5 mJ/m$^2$ in the micromagnetic approximation for their system. Since the DMI originates from the interface between the heavy metal substrate and the ferromagnet, the corresponding strength of $D$ for the parameters used here and in Thiaville \emph{et al}.~\cite{Thiaville:2012ia} with the same atomistic interaction are 2 mJ/m$^2$ and 3.3 mJ/m$^2$, respectively, as a result of the larger ferromagnetic film thicknesses. While the figure of 2 mJ/m$^2$  is smaller than range of parameters studied here, the general features reported for the breathing dynamics are expected to remain valid, as shown in Fig.~\ref{fig:freq_vs_Bz_BC}.

It could be argued that the dynamic susceptibility, $\chi_{zz}(\omega)$, would also be finite with the presence of a magnetic bubble state, which could be stable under dipolar fields alone without the need for the DMI. A magnetic bubble comprises a region of reversed magnetization and can share the same topology as the skyrmion state considered, i.e., bubbles possess the same integer topological charge $Q = (1/4\pi) \int dx dy \; \mathbf{m} \cdot \left( \partial_x \mathbf{m} \times \partial_y \mathbf{m}  \right)$ as skyrmions, but they are not compact and their size is determined primarily by the dipolar interaction. In the dot geometry studied, it has been verified that magnetic bubbles are not stable with the parameter sets considered if $D = 0$. The inclusion of a large density of pinning sites, related to material defects or film roughness, might be sufficient to pin magnetic bubbles in the dot geometry considered, but detailed investigations along this line are beyond the scope of the present study. In the defect-free system at least, it is our contention that the low-frequency breathing mode and the finite susceptibility $\chi_{zz}(\omega)$ for perpendicular driving fields are robust signatures for the presence of individual skyrmions in submicron dots.

The gyrotropic dynamics of the skyrmion core has not been examined here, as discussed earlier. The perpendicular driving field has been used precisely to minimize such dynamics. In experiment, however, thermal fluctuations could lead to low-amplitude gyrotropic motion about the dot center, much in the same way that vortex gyration in dots can be induced thermally.~\cite{Wysin:2012km} However, such gyration frequencies are typically in the sub-GHz regime,~\cite{Moutafis:2009jl, Makhfudz:2012eh, Dai:2013eb} which are an order of magnitude lower than the breathing mode frequencies computed here. Given this difference, small amplitude gyration is not expected to influence the breathing properties in any significant way.

In summary, the breathing modes of single skyrmion states in circular ultrathin dots have been studied using micromagnetics simulations and how these modes couple to the spin wave eigenmodes in this geometry have been examined. Distinct features in the power spectra of these modes in perpendicular static and dynamic magnetic fields allow such skyrmion states to be detected in experiment. These features are shown to be robust for different material systems.

\section{Acknowledgements}
We thank Noah Vanhorne and Nicolas Reyren for stimulating discussions. This work was partially supported by the French National Research Agency (ANR) under contract no. ANR-11-BS10-0003 (NanoSWITI).

\bibliography{articles}

\end{document}